\documentclass[prl,nofootinbib,twocolumn,superscriptaddress]{revtex4}
\def\mysection#1{{\bf #1.} }

\usepackage{amssymb}
\usepackage{amsmath}

\usepackage[dvips]{graphicx}
\usepackage{longtable}
\usepackage{verbatim}
\usepackage{amsfonts}

\newcommand{\beq}{\begin{equation}}
\newcommand{\eeq}{\end{equation}}
\newcommand{\beqa}{\begin{eqnarray}}
\newcommand{\eeqa}{\end{eqnarray}}
\newcommand{\no}{\nonumber}

\newcommand{\ket}[1]{\left\vert #1\right\rangle}

\newcommand{\lsim}{\mathrel{\rlap{\lower4pt\hbox{\hskip1pt$\sim$}}
    \raise1pt\hbox{$<$}}}         
\newcommand{\gsim}{\mathrel{\rlap{\lower4pt\hbox{\hskip1pt$\sim$}}
    \raise1pt\hbox{$>$}}}         
\begin{document}

\preprint{{\vbox{\hbox{}\hbox{}\hbox{}
\hbox{hep-ph/yymmnnn}}}}
\vspace*{0.5cm}
\title{The importance of $N_2$ leptogenesis}

\author{Guy Engelhard}
\affiliation{Department of Particle Physics,
  Weizmann Institute of Science, Rehovot 76100, Israel}
\author{Yuval Grossman}
\affiliation{Department of Physics, Technion-Israel Institute of
  Technology, Technion City, Haifa 32000, Israel}
\author{Enrico Nardi}
    \affiliation{INFN, Laboratori Nazionali di Frascati, C.P. 13,
      100044 Frascati, Italy}
    \affiliation{Instituto de F\'{i}sica, Universidad de Antioquia,
    A.A.1226, Medell\'{i}n, Colombia}
\author{Yosef Nir}
\affiliation{Department of Particle Physics,
  Weizmann Institute of Science, Rehovot 76100, Israel}
\date{\today}

\begin{abstract}
  We argue that fast interactions of the lightest singlet neutrino
  $N_1$ would project part of a preexisting lepton asymmetry $L_p$
  onto a direction that is protected from $N_1$ washout effects, thus
  preventing it from being erased. In particular, we consider an
  asymmetry generated in $N_2$ decays, assuming that $N_1$
  interactions are fast enough to bring $N_1$ into full thermal
  equilibrium. If $N_1$ decays occur at $T\gsim 10^9\,$GeV, that is,
  before the muon Yukawa interactions enter into thermal equilibrium,
  then generically part of $L_p$ survives. In this case some of the
  constraints implied by the standard $N_1$ leptogenesis scenario hold
  only if $L_p\approx0$.  For $T\lsim 10^9\,$GeV, $L_p$ is generally
  erased, unless special alignment/orthogonality conditions in flavor
  space are realized.
\end{abstract}

\maketitle

\mysection{Introduction}
\label{sec:introduction}
The existence of heavy singlet neutrinos with Majorana masses is quite
generic in theories that go beyond the Standard Model (SM) and,
furthermore, can simultaneously explain two puzzles: It can induce,
via the see-saw mechanism,  
active neutrino masses that are much lighter than those of the SM charged
fermions, and it can generate, via leptogenesis \cite{Fukugita:1986hr}, the
baryon asymmetry of the Universe (BAU).

Leptogenesis is of major phenomenological interest because, within the
conventional leptogenesis scenario, the measured value of the BAU has
non-trivial implications for other fundamental issues. In particular,
it provides constraints on low energy neutrino parameters, on the mass
scale of the heavy Majorana neutrinos, and on the reheating
temperature after inflation. In order to understand how robust are
such implications, it is important to explore under which
conditions the conventional picture holds. The aim of this paper is to
show that, even within the standard scenario, one of the assumptions that
underlies most of the leptogenesis analyses might not hold, implying
that the standard leptogenesis constraints are relaxed or even evaded.

In the conventional picture, three singlet neutrinos $N_\alpha$ are
added to the SM, with hierarchical Majorana masses $M_1 \ll M_2 \ll
M_3$. It is often assumed that the $L$-violating interactions of $N_1$
would washout any lepton asymmetry $L_p$ generated at temperatures
$T\gg M_1$ and, in particular, the asymmetry generated in the decays
of $N_{2,3}$. If this were the case, the final asymmetry would depend
only on $N_1$ dynamics, and the number of parameters relevant for
leptogenesis would be reduced to just a few: the mass of the lightest
singlet neutrino $M_1$, the ``washout parameter'' $\tilde m_1$
(defined below), and the $CP$ asymmetry in $N_1$ decays
$\epsilon_{N_1}$. We show that, contrary to common wisdom, part of
the lepton asymmetry generated in $N_{2,3}$ decays in general survives
the $N_1$ leptogenesis phase.  Thus, it is quite possible that the
lepton asymmetry relevant for baryogenesis originates mainly from
$N_{2,3}$ decays.
The possibility that $N_2$ leptogenesis could successfully explain the
BAU has been already suggested in some recent papers
\cite{Vives:2005ra,DiBari:2005st,Blanchet:2006dq}.  However, the
scenario put forth in these analyses is essentially that of
``$N_1$-decoupling'', in which the Yukawa couplings of $N_1$ are
simply too weak to washout the $N_2$-generated asymmetry. Here we
point out that $N_1$ decoupling is not a necessary condition, and that
$N_2$ leptogenesis can be successful even when the Yukawa interactions
of $N_1$ are fast enough to bring these states into complete thermal
equilibrium.
Our main conclusions, regarding the survival of part of the asymmetry
generated in $N_2$ decays, have been previously stated in refs.
\cite{Barbieri:1999ma,Strumia:2006qk}. We describe the framework where this
picture arises, spelling out the assumptions involved and uncovering some
important subtleties, and we explain the importance of the results for the
analysis of leptogenesis.

\noindent
\mysection{Notations}
\label{sec:not-and-for}
The particles that play a role in leptogenesis are the heavy singlet
neutrinos $N_\alpha$, the light lepton $SU(2)$-doublets $L_i$ and
$SU(2)$-singlets $E_i$, and the standard model Higgs $H$. The
relevant Lagrangian terms are 
\begin{equation}
  \label{eq:1}
-\mathcal{L} =\frac{1}{2} M_\alpha N_\alpha N_\alpha + 
\lambda_{\alpha i}HN_\alpha L_i + Y_i
 H^\dagger L_i E_i. 
\end{equation}
where $\alpha=1,2,3$ is a heavy neutrino index while $i=e,\mu,\tau$ is a
flavor index.  Eq.~(\ref{eq:1}) is written in the mass basis for the singlet
neutrinos and for the charged leptons, that is, $M$ and $Y$ are
diagonal. It is convenient to define $\ell_\alpha$, the three (in
general non-orthogonal) combinations of lepton doublets to which the
corresponding $N_\alpha$ decay:  
\beq
\ket{\ell_\alpha}=(\lambda\lambda^\dagger)_{\alpha\alpha}^{-1/2}
\sum_i\lambda_{\alpha i}\ket{L_i}.
\eeq
We denote by $n_{\ell_\alpha}$ the number density of $\ell_\alpha$ and
normalize the asymmetry-density to the equilibrium density $n_{\rm eq}$:
\beq
y_{\ell_\alpha}=(n_{\ell_\alpha}-n_{\overline{\ell}_\alpha})/n_{\rm eq}.
\eeq
We use similar notations for other particle species, {\it
  e.g.} $n_H$ and $y_H$ for the Higgs.

The CP asymmetry generated in $N_\alpha$ decays reads:
\begin{equation}\label{eq:eps}
\epsilon_{N_\alpha}=\frac{\Gamma(N_\alpha\to\ell H)-\Gamma(N_\alpha\to\bar\ell
  \bar H)}{\Gamma(N_\alpha\to\ell H)+\Gamma(N_\alpha\to\bar\ell \bar H)}.
\end{equation}
We introduce  three quantities $\tilde m_\alpha$ with dimension of a mass that
parametrize the washout of the lepton asymmetry $\epsilon_{N_\alpha}$ 
due to $N_\alpha$-related interactions:
\begin{equation}
  \label{eq:tmi}
  \tilde m_\alpha = \frac{\langle H\rangle^2 
(\lambda\lambda^{\dag})_{\alpha\alpha}}{M_\alpha}.
\end{equation}
The baryon number generated from the decays of the $N_{1,2,3}$ neutrinos,
normalized to the entropy density $s$ can be formally written as follows:
\begin{equation}
  \label{eq:2}
  Y_{\mathcal{B}}\equiv 
\frac{n_B-n_{\overline{B}}}{s} = -1.38\times 10^{-3}\sum_{\alpha,\beta}
\epsilon_{N_\alpha}\eta_{\alpha\beta},
\end{equation}
where $\eta_{\alpha\beta}$ are efficiency factors that are related to
the effects of $N_\beta$ interactions on the asymmetry
$\epsilon_{N_\alpha}$~\cite{footnote1}.  The diagonal factors
$\eta_{\alpha\alpha}$ are simply related to washout effects. The
off-diagonal factors $\eta_{\alpha\beta}$ ($\alpha\neq \beta$) have a
more complicated structure since, as we discuss below, they represent
also decoherence effects on $\ell_\alpha$ induced by $N_\beta$
interactions. 
In the standard leptogenesis scenario (and in the one-flavor
approximation), only one efficiency factor is relevant, $\eta_{11}$.
Calculating this factor requires solving the complete set of Boltzmann
equations.  However, for $M_1\ll 10^{14}$ GeV (when $\Delta L=2$ washouts
can be neglected) $\eta_{11}$ is completely determined by the
mass and 
couplings of $N_1$ via the combination $\tilde m_1$ defined in
Eq.~(\ref{eq:tmi}).  A simple relation that approximates quite well the
dependence of $\eta_{11}$ on $\tilde m_1$ can be found {\it e.g.}  in
\cite{Giudice:2003jh}.

Recent data on the cosmic background anisotropy \cite{Spergel:2006hy} and
considerations of Big Bang Nucleosynthesis \cite{Steigman:2005uz} yield
\begin{equation}
  \label{eq:4}
 Y_{\mathcal{B}}^{\rm exp}= (8.7\pm 0.3)\times 10^{-11}.
\end{equation}
This is the experimental number that leptogenesis aims to explain.  In the
following  we discuss the possibility that this number
could be related to $\epsilon_{N_2}$, rather than to~$\epsilon_{N_1}$.

\mysection{The framework} 
We are interested in the strong washout regime for $N_1$-related
interactions:
\beq\label{largetm} 
\tilde m_1\gg m_*,
\eeq
where $m_*=1.66 \sqrt{g_*} 16\pi \langle H\rangle^2/M_P\approx 10^{-3}\,$eV, $g_*$ is the
number of degrees of freedom (d.o.f.) in equilibrium and $M_P$ is the
Plank mass.  In the weak washout regime ($\tilde m_1\lsim m_*$) and in
the $N_1$ decoupling regime ($\tilde m_1 \ll m_*$) the final
$B-L$ asymmetry depends in general on the initial conditions for $N_1$
leptogenesis.  We will find that, contrary to common belief,
this is true also for the regime of strong $N_1$ washouts.  

We consider a lepton asymmetry $L_p\neq 0$ that is present before the
onset of $N_1$ leptogenesis. For definiteness we study the case where
$L_p$ originates from out-of-equilibrium, $CP$ violating decays of
$N_2$.  (The effects of $N_3$ decays can be readily included in the
same way as $N_2$ effects, and should be taken into account in
quantitative estimates of the BAU resulting from leptogenesis.) 
We assume that $N_2$-related washouts are not too strong: 
\beq\label{smalltm} 
\tilde m_2\not\gg m_*,
\eeq
so that a finite $L_p$ survives \cite{nodeg}.

Our analysis is qualitative. To simplify our arguments, we impose two
further conditions: thermal leptogenesis, that is $n_{N_2}(T\gg M_2)\approx0$
and in particular
\beq\label{thermallg} n_{N_1}(T\gg M_1)\approx0, 
\eeq
and strong hierarchy,
\beq\label{strhie}
M_2/M_1\gg1.
\eeq

We now describe the simplifications that occur when the conditions
(\ref{largetm})-(\ref{strhie}) are realized.  The hierarchy
(\ref{strhie}) is assumed to be large enough that, together with
Eq. (\ref{thermallg}), $n_{N_1}(T\sim M_2)\approx 0$ is guaranteed. On the
other hand, the Boltzmann suppression gives $n_{N_2}(T\sim M_1)\approx 0$. This
situation has several consequences: {\it (a)} there are no $N_1$
related washout effects during the $N_2$ leptogenesis, {\it (b)} there
are no $N_2$ related washout effects during $N_1$ leptogenesis, {\it
(c)} we do not need a density matrix to describe the $\ell_1$ and $\ell_2$
densities during $N_2$ leptogenesis, and {\it (d)} the density matrix
for $\ell_1$ and for the states orthogonal to $\ell_1$ does not depend on
the  interactions with $N_2$.

Condition (\ref{largetm}) implies that, already at $T\gsim M_1$,
the interactions mediated by the $N_1$ Yukawa couplings become
sufficiently fast to quickly destroy the coherence of the state $\ell_2$
produced in $N_2$ decays.  Then a statistical mixture of $\ell_1$ and of
the states orthogonal to $\ell_1$ builds up, and it can be described by a
{\it diagonal} density matrix.  The analogous situation in
which decoherence occurs because of fast charged lepton Yukawa
interactions was analyzed
in~\cite{Barbieri:1999ma,Endoh:2003mz,Abada:2006fw,Nardi:2006fx,%
DeSimone:2006dd} in the context of flavored leptogenesis.  It was
found \cite{Abada:2006fw,DeSimone:2006dd} that after a short transient
period during which the flavor-components of the asymmetry undergo
fast oscillations, they decohere and get projected onto the flavor
basis.  There is, however, a difference between the present case and
the flavor case: as the temperature drops below $M_1$, the $N_1$
decohering interactions are suppressed by $T^2/M_1^2$ and thus slow down with
respect to the Universe expansion rate.  When $T/M_1 < m_* /
\tilde m_1$, they become irrelevant. Our assumption (\ref{largetm})
implies that this happens only at $T\ll M_1$, that is after quantum
coherence between the $\ell_2$ components has been completely destroyed.
Finally, it is crucial that at $T\sim M_1$ the effects related to $N_2$
interactions are suppressed enough not to interfere with the
decoherence effects of  $N_1$, and this is ensured if
the condition $M_1/M_2 < m_* / \tilde m_2$ holds. The combination of
Eq. (\ref{smalltm}) with Eq. (\ref{strhie}) implies that this
condition is indeed fulfilled.

On general grounds one would expect that decoherence effects proceed
faster than washout. In the relevant range, $T\gsim M_1$, this is also
ensured by the fact that the dominant ${\cal O}(\lambda^2)$ washout
processes, that is the inverse decays $\ell H \to N_1$, are blocked because
of thermal effects \cite{Giudice:2003jh}, and only scatterings with
top-quarks and gauge bosons, that have additional suppression factors
of $Y_t^2$ and $g^2$, contribute to the washout.

In the following we analyze different temperature regimes, paying
attention to the cases where the interplay between flavor effects and
decoherence effects induced by the $N_1$ Yukawa couplings becomes
important.

 \smallskip

\mysection{$\mathbf{M_{1,2}\gsim10^{12}}$ GeV}
\label{sec:unflavored}
We first consider the case where both $N_{2}$ and $N_1$ decay at
$T\gsim10^{12}\,$GeV, {\it i.e.} when all charged lepton Yukawa interactions
are slower than the expansion rate of the Universe and flavor effects are
irrelevant.  (Note that $\Delta L=2$ interactions of the heavier neutrinos can
generally be in equilibrium down to $T\sim 10^{13}\,$GeV. Taking
these interactions into account complicates the analysis
considerably. We thus neglect these interactions, keeping in mind that
the range of temperatures and parameters where our analysis applies  
straightforwardly is rather restricted.) 

During the stage of $N_2$ decays we have, as explained above, $n_{N_1}\approx
0$. Consequently, there are no $N_1$-related washout effects during $N_2$
leptogenesis.  Since we assume $\eta_{22}\not\ll1$, sizeable asymmetries can
survive the $N_2$-related washouts and, since we neglect $N_3$-related
effects, we denote the total lepton asymmetry (corresponding to the trace of
density matrix of the asymmetries) present at the end of $N_2$
leptogenesis by $y_{\ell_2}$. We thus have
\begin{eqnarray} 
\label{eq:phase1a} 
y_{\ell_2} &\sim& \eta_{22} \epsilon_{N_2}. 
\end{eqnarray} 
Decoherence effects induced by the $N_1$ Yukawa interactions 
become important at a later stage. For $\tilde m_1$ large enough (8),
they become faster than the expansion rate of the Universe already at
$T\gsim M_1$.  To study the fate of $y_{\ell_2}$ through the $N_1$
leptogenesis phase, it is convenient to choose the (orthogonal) basis
$(\ell_1,\,\ell_0,\,\ell'_0)$ where, without loss of generality,
$\langle\ell'_0|\ell_2\rangle=0$. Due to the invariance of the trace
under a change of basis, the lepton asymmetry $y_{\ell_2}$ of
Eq.~(\ref{eq:phase1a}) decomposes into the two components: 
\begin{equation}
\label{eq:c2}
y_{\ell_2}=y_{\ell_0}+y_{\ell_1}\,.  
\end{equation}
In general (and, in particular, if there is no fine-tuned alignment of
$\ell_1$ and $\ell_2$) we expect $y_{\ell_0}={\cal O}(y_{\ell_2})$.
The crucial point is that since $\ell_0$ is orthogonal to $\ell_1$,
$y_{\ell_0}$ is protected against $N_1$ washouts. This ensures that a
finite part of the asymmetry generated in $N_2$ decays survives the
$N_1$ leptogenesis phase, and is a source of $Y_L\neq0$.

Actually, by taking into account the constraints on the various
chemical potentials implied by the reactions that are in equilibrium
in this temperature range, one can get a better insight about the true
composition of the surviving lepton asymmetry.  To simplify the
analysis let us assume that $\tilde m_1$ is so large that around $T\sim
M_1$, when the number density of $N_1$ is maximal, $N_1$-interactions
attain chemical equilibrium. This implies:
\beq \label{eq:yN1}
y_{N_1}=y_{\ell_1}+y_H/2,
\eeq
where $y_{N_1}$ corresponds to a chemical potential for $N_1$ that we
will eventually set to zero.  Combining (\ref{eq:yN1}) with the
conditions implied by in-equilibrium gauge and top-Yukawa interactions
and by baryon number and hypercharge conservation, we obtain
\beqa
y_H&=&(2/3)(y_{\ell_1}+y_{\ell_0}),\no\\
y_{N_1}&=&(4/3)y_{\ell_1}+(1/3)y_{\ell_0},\no\\
Y_L/Y^{\rm eq}&=&(10/3)y_{\ell_1}+(7/3)y_{\ell_0}, 
\eeqa 
where $Y^{\rm eq}$ is the equilibrium density of relativistic species
with one d.o.f..  At temperatures $T\sim M_1$ the $L$-violating
interactions of $N_1$ are fast and imply $y_{N_1}\to0$, finally yielding
$y_H=y_{\ell_0}/2$ and $y_{\ell_1}=-y_{\ell_0}/4$. We note that since the
strong washout (equilibrium) condition (\ref{eq:yN1}) does not imply
$y_{\ell_1}=0$ and $y_H=0$ separately, part of the asymmetry in $\ell_1$,
that is a quarter in size and opposite in sign to $y_{\ell_0}$, also
survives, yielding a total lepton asymmetry that is somewhat smaller
than $2y_{\ell_0}$:
\beq 
Y_L/Y^{\rm eq}=(3/2)y_{\ell_0}.  
\eeq
In a fine-tuned situation where $\ell_2\parallel \ell_1$ and
$y_{\ell_0}=0$, $y_{\ell_2}$ can be completely washed out by $N_1$
interactions.  (This alignment condition would entail one massless
neutrino.)  In this case, however, some fraction of $y_{\ell_3}$
generated in $N_3$ decays remains protected from $N_{1,2}$ washouts.
One cannot further impose $\ell_3\parallel \ell_2\parallel \ell_1$
since this would imply two massless neutrinos.

\smallskip 
\mysection{$\mathbf{10^9\ {\rm GeV}\lsim M_1\lsim 10^{12}\ {\rm GeV}\lsim M_2}$}
\label{sec:twoflavors}
We now consider the case where {\it (a)} $N_1$ decays at $10^9\ {\rm
  GeV}\lsim T\lsim10^{12}\,$GeV, when the $\tau$ (but not the $\mu$)
Yukawa interactions are in equilibrium, while {\it (b)} $N_2$
leptogenesis occurs in the ``flavor blind'' regime, $T\gsim
10^{12}\,$GeV yielding the result in Eq.~(\ref{eq:phase1a}).  To study
the onset of fast $N_1$ interactions, we use the orthogonal basis
$(\ell_\tau,\ell_a,\ell_0)$ where, without loss of generality,
$\langle\ell_0|\ell_1\rangle=0$ while (except for the specific cases
where $\ell_\tau$ is either aligned with or orthogonal to $\ell_1$)
$\langle\ell_{\tau,a}|\ell_1\rangle\neq 0$.
Given that $\ell_0$ is affected by neither $\tau$ nor $N_1$
interactions, its off-diagonal quantum correlations are quickly
damped, singling out the $\ell_0$ component of $\ell_2$. The crucial
quantity is then $y_{\ell_0}$, the projection of the lepton asymmetry
onto the $\ell_0$ direction.  We can again conclude that this part of
the lepton asymmetry remains protected against $N_1$ washouts.

Compared to the previous case, however, we now have a larger number of
interactions that are in equilibrium. As a representative set for this
temperature interval, we take the interactions of the third family
Yukawas $Y_{t,b,\tau}$ and the QCD and electroweak sphalerons.  The
latter violate baryon number, but conserve the three charges $\Delta_i\equiv
B/3-L_i$ (for the case at hand $i=\tau,a,0$).  Furthermore, any initial
asymmetry in $\Delta_{0}$ generated during $N_2$ decays
is conserved also by $N_1$ interactions.  After solving for the
equilibrium constraints and setting $y_{N_1}=0$ we find:
\begin{eqnarray}
\label{eq:B-L}
Y_{B-L}&=&+(90/107)\,Y_{\Delta_{0}}, \\ 
\label{eq:tau}
Y_{\Delta_{\tau}}&=&-(16/107)\,Y_{\Delta_{0}},\\  
\label{eq:ella}
Y_{\Delta_{a}} &=& -(1/107) Y_{\Delta_{0}},  
\end{eqnarray}
where $Y_{\Delta_{0}}\equiv -2 y_{\ell_0} Y^{\rm eq}$. We see that, in
spite of having equilibrium reactions with $N_1$, fractions of
$y_{\ell{\tau}}$ and $y_{\ell_a}$ survive. In addition, the fast
$\tau$ Yukawa interactions generate $y_\tau$, and the fast sphalerons
generate a net baryon number. The overall effect is to cancel a
(small) part of the $y_{\ell_0}$ contribution to the final $B-L$.

In deriving Eqs.~(\ref{eq:B-L})-(\ref{eq:ella}) we assume that $\ell_1$ has
sizable projections on both the $\ell_{\tau,a}$ states, leaving aside the fine
tuned possibilities $\ell_1\perp \ell_\tau(\ell_a)$.  Thus we apply the two
equilibrium conditions $y_{\ell_\tau}=y_{\ell_a}=-y_H/2$.  In the realistic
case of large but not unnaturally large $N_1$ washouts, the condition
$y_{\ell_\tau}=y_{\ell_a}$ is not expected to hold, since the projections of
$y_{\ell_2}$ onto these two states, as well as their specific washouts,
differ. Solving for the coupled evolution of $y_{\ell_\tau}$ and $y_{\ell_a}$
is a non-trivial problem, given that their off-diagonal correlations, that are
maintained by the $N_1$ Yukawa interactions, are likely to be damped only at
small $T$, when the $N_1$ effects are suppressed by $T^2/M_1^2$.  
The detailed evolution of the $\ell_\tau$-$\ell_a$ system represents an
interesting quantitative problem, but will not change the main qualitative
result: Part of the $y_{\ell_2}$ asymmetry survives the $N_1$ leptogenesis
phase.  If also $N_2$ leptogenesis occurs at $T\lsim10^{12}\,$GeV, some
details of the previous analysis change, but the final conclusion remains. We
conclude that if leptogenesis occurs in the theoretically preferred range
$T\gsim10^{9}\,$GeV, when the $\mu$ Yukawa reactions are still slow, the
constraints inferred by requiring successful $N_1$ leptogenesis hold only
under the additional assumption that either $\epsilon_{N_2}$ is small enough
or $N_2$-related washout is strong enough ($\eta_{22} \ll 1$) to drive
$y_{\ell_2}\to 0$.

\smallskip
\mysection{$\mathbf{M_1\lsim 10^9\ {\rm GeV}}$}
\label{sec:threeflavors}
The last possibility is that $N_1$ decays at $T\lsim 10^9\,$GeV (the
$N_2$ decay temperature is irrelevant here).  In this regime, because
of the fast $\tau$ and $\mu$ Yukawa reactions, the full lepton flavor
basis $(\ell_e,\ell_\mu,\ell_\tau)$ is resolved.  In the general
situation, where $\langle\ell_i \ket{\ell_1} \not\ll 1$ for
$i=e,\,\mu,\,\tau$, there are no directions in flavor space where an
asymmetry remains protected from $N_1$-related washouts, and $N_2$
leptogenesis fails.  A preexisting asymmetry can only survive if it is
stored in a flavor that is weakly coupled to $N_1$.

\smallskip
\mysection{Conclusions} 
\label{sec:conclusions} 
The effect of $N_1$ interactions on $N_2$ leptogenesis is twofold:
decoherence of $\ell_2$, and washout of the lepton-asymmetry $y_{\ell_2}$
generated in $N_2$ decays.  For weak $N_1$ washouts ($\tilde
m_1\ll10^{-3}$ eV) neither effect is important, and $y_{\ell_2}$ survives.
For $N_1$ washouts in the strong regime ($\tilde m_1\gg10^{-3}$ eV),
both effects are important. For a generic flavor structure,
$N_1$-related decoherence effects project part of $y_{\ell_2}$ onto a
flavor direction that is protected against $N_1$ washout, and this
component of the initial asymmetry survives.  Previous analyses
ignored $N_1$-related decoherence effects, and assumed that when $N_1$
leptogenesis occurs in the strong washout regime the final BAU is
independent of initial conditions.  We find that this assumption is
justified only in the following cases: {\it i)}~Vanishing decay
asymmetries and/or efficiency factors for $N_{2,3}$ ($\epsilon_{N_{2}}\eta_{22}\approx
0$ and $\epsilon_{N_{3}}\eta_{33}\approx 0$); {\it ii)}~$N_1$-related
washouts are still significant at $T\lsim10^9\,$GeV; {\it
iii)}~Reheating occurs at a temperature in between $M_2$ and $M_1$.
In all other cases the $N_{2,3}$-related parameters cannot be ignored
when calculating the BAU.

\smallskip\noindent
{\bf Acknowledgments.}
We thank Michael Dine, Marta Losada, Michael Peskin, Antonio Riotto, Esteban
Roulet, and Adam Schwimmer for useful discussions.  This project was supported
by the Albert Einstein Minerva Center for Theoretical Physics. The work of
Y.G. is supported in part by the Israel Science Foundation under Grant No.
378/05. The work of E.N. is supported in part
by Colciencias 
under contract 1115-333-18739.  The research of Y.N.  is supported by the
Israel Science Foundation founded by the Israel Academy of Sciences and
Humanities, and by a grant from the United States-Israel Binational Science
Foundation (BSF), Jerusalem, Israel.  E.N. and Y.N. thank the organizers of
the International Workshop on High Energy Physics in the LHC Era held in
Valparaiso-Chile where this work was completed, for their kind invitation and
pleasant working atmosphere.



\begin{thebibliography}{99}
\bibitem{nodeg}
It is impossible to satisfy Eq. (\ref{smalltm}) for the case of
quasi-degenerate neutrinos. For this (and other) reasons, we
implicitly assume hierarchichal light neutrinos masses.
\end{thebibliography}
\end{document}